\documentclass[twocolumn,showpacs,preprintnumbers,amsmath,amssymb,prl,superscriptaddress]{revtex4-1}
\usepackage{graphicx}
\usepackage{dcolumn}
\newcommand{\tc}[1]{\multicolumn{1}{c}{#1}} 
\newcommand{\tr}[1]{\multicolumn{1}{r}{#1}} 

\usepackage{booktabs}
\usepackage{bm}
\newcolumntype{.}[1]{D{.}{.}{#1}}
\usepackage{float}
\usepackage{color}
\usepackage{hyperref}
\hypersetup{
	breaklinks=true,
	pdfnewwindow=true,      
	colorlinks=true,       
	linkcolor=blue,          
	citecolor=blue,        
	filecolor=blue,      
	urlcolor=blue,          
}
\begin{document}
\title{High-pressure lithium as an elemental topological semimetal}
\author{S. F. Elatresh}
\affiliation{Department of Chemistry and Chemical Biology, Cornell University, Baker Laboratory, Ithaca, New York 14853-1301, USA}
\author{Zhimou Zhou}
\affiliation{International Center for Quantum Materials, School of Physics, Peking University, Beijing 100871, China}
\author{N. W. Ashcroft}
\affiliation{Laboratory of Atomic and Solid State Physics, Cornell University, Ithaca, New York 14853-1301, USA}
\author{S. A. Bonev}
\email[Electronic address:] { bonev@llnl.gov, jfeng11@pku.edu.cn,
  rh34@cornell.edu}
\affiliation{Lawrence Livermore National Laboratory, Livermore, California 94550, USA}
\author{Ji Feng}
\email[Electronic address:] { bonev@llnl.gov, jfeng11@pku.edu.cn,
  rh34@cornell.edu}
\affiliation{International Center for Quantum Materials, School of
  Physics, Peking University, Beijing 100871, China}
\affiliation{Collaborative Innovation Center of Quantum Matter,
  Beijing 100871, China}
\affiliation{CAS Center for Excellence in Topological Quantum Computation, University of Chinese Academy of Sciences, Beijing 100190, China}
\author{Roald Hoffmann}
\email[Electronic address:] { bonev@llnl.gov, jfeng11@pku.edu.cn,
  rh34@cornell.edu}
\affiliation{Department of Chemistry and Chemical Biology, Cornell University, Baker Laboratory, Ithaca, New York 14853-1301, USA}

\begin{abstract}
 Topological semimetals generally contain heavy elements. Using density-functional theoretic calculations,
we predict that three dense lithium polymorphs in the pressure range 200--360 GPa
display nontrivial semimetallic electronic structure. Specifically, these 
high-pressure phases exhibit Fermi pockets which are degenerate over a loop in $\boldsymbol{k}$-space, around which an encircling $\bm k$-space path is threaded by $\pm \pi$ Berry phase. Accordingly, these dense lithium phases are topological nodal loop semimetals involving a single light element. 

 \end{abstract}

\date{\today}
\maketitle

\section{Introduction}
Pressure can induce dramatic changes to a material's geometric and
electronic structures, resulting in highly complex chemical and
physical behavior. At ambient pressure, the light alkali metals, such
as lithium and sodium, are viewed as simple metals; the nearly-free
electron approximation picture adequately describes their valence
electrons. However, the phase diagrams of Li and Na are riddled with  novel
features, which are clearly non-simple~\cite{PhysRevB.39.10552,Neaton1999,PhysRevLett.94.185502,Tuoriniemi2007,Raty:2007iz,PhysRevLett.101.075703,Matsuoka:2009tf,Ma2009TransparentDS,Lazicki:2010wd,Guillaume2011,PhysRevLett.108.055501,Elatresh23052017}.
Lithium, in particular, patently departs from the realm of simple metals under the
application of external pressure

At ambient pressure and temperature, lithium is a good metal
forming a simple and highly symmetric body centered cubic ({bcc})
crystal structure. As the pressure is increased, starting 
near 40~GPa, it undergoes  several transformations to complex low-symmetry phases.
The structural changes are coupled to counterintuitive modifications
of its electronic properties, which cannot be explained by a theory
based on the nearly-free-electron approximation.  A number of exotic
high pressure phenomena ensue, including a metal to
semiconductor transition, reappearance of metallization, superconductivity,  anomalous
melting curve, and electride properties 
~\cite{Neaton1999,Guillaume2011,Hanfland2000, Matsuoka:2009tf,PhysRevLett.102.146401,PhysRevLett.109.185702, PhysRevLett.101.075703, PhysRevB.94.104107,PhysRevB.89.144103,doi:10.1063/1.4928076,Yu2018}.

Clearly, these changes must be accompanied by
nontrivial electronic band structure modifications.
Most of the previous studies on the electronic properties of
dense solid lithium have
focused on the metal-to-semiconductor transition found around 80~GPa
pressure~\cite{PhysRevB.89.144103} and on its superconducting properties from 
35 to 48~GPa~\cite{PhysRevLett.86.1861,Shimizu2002,Ashcroft2002,Struzhkin1213,1742-6596-121-5-052003,PhysRevB.82.184509}
 and beyond~\cite {C8CP05455J}. At higher pressures and low temperature,
the research to date has mainly focused on determining the preferred  Li crystal
structures~\cite{PhysRevLett.102.146401,PhysRevLett.106.015503}.
It has been shown that the high-pressure phases
are characterized with the appearance in the electron band structure of
a pseudogap, $s$-to-$p$
charge transfer, 
and electride-like localization of the conduction 
electrons in interstitial ionic 
regions~\cite{Rousseau2011,doi:10.1021/ar4002922,PhysRevLett.102.146401,PhysRevLett.106.015503,Yu2018}. However, the
possibility for re-appearing semiconducting behavior or the emergence of
states with non-trivial topology has
not been examined in detail.

Topologically non-trivial metals and semimetals have been
the focus of much interest in modern condensed matter physics
research. Their electronic structures exhibit nontrivial band
crossings near the Fermi energy, around which the low-energy
excitations behave differently from the conventional
Schr\"{o}dinger-type fermions. For example, Weyl and Dirac semimetals
host isolated twofold and fourfold degenerate points, respectively,
with linear energy dispersions. Their  electronic excitations are
analogous to the relativistic Weyl and Dirac
fermions~\cite{Wan11prb,Wang12prb,Wang13prb,Liu14sci,Xu15sci}, making
it possible to simulate interesting high-energy physics phenomena in
condensed matter systems~\cite{Guan17npj}. Most materials studied either consist of binary compounds of  
heavy elements, such
as Cd$_3$As$_2$~\cite{Wang13prb}, Na$_3$Bi~\cite{Wang12prb,Liu14sci},
or are still more complex compounds~\cite{Li18prb}. 
Although some Dirac-like features have been observed in the band
  structures of lithium subhydrides under
  pressure~\cite{Hooper12cpc} and predicted for dense hydrogen surface states~\cite{Naumov16prl} , topological semimetals involving only a single light element are hitherto unknown~\cite{Xu17}.

In this work, we report a first-principles study on the lightest metallic element, lithium, 
under high pressure ranging from 200 to 360~GPa. We focus on the three stable phases
 in this pressure range, namely those with space groups {\it Cmca}-24, {\it Cmca}-56, and  {\it P}4$_2$/{\it mbc}. 
We find that in all these phases, high pressure tends to drive the
valence electrons to localize in the interstitials of the rather
complex Li networks. We also show that their electronic structures
share some common features. Importantly, all of these three phases turn out to be topological nodal-loop semimetals.
If the predictions for dense lithium made here are confirmed by
 experiments, lithium would be the simplest and lightest three-dimensional
(3D) topological nodal-loop semimetal material discovered.

\section{Computational Methods}

Full structural optimization, enthalpy, 
electronic band structure, 
and density of states 
calculations
were performed within density-functional theory (DFT)  
using the  ABINIT code~\cite{Gonze20092582} 
with three-electron Hartwigsen-Goedeker-Hutter pseudopotential~\cite{PhysRevB.58.3641}, 
and the generalized gradient approximation parametrized 
by Perdew, Burke, and Ernzerhof (GGA-PBE)~\cite{prl96Perdew}.
A plane-wave expansion with a 2700 eV cut-off and a {\bf k}-point 
grid for self-consistent calculations as large
 as $10\times10\times10$, $6\times6\times6$,  and $16\times16\times16$
 were used for  {\it Cmca}-24,
  {\it Cmca}-56, and  {\it P}4$_2$/{\it mbc}, respectively. These
  dense {\bf k}-point grids  are sufficient to  
  ensure convergence for enthalpies to better than 1~meV/atom.

For the rest of the analysis, 
DFT calculations were performed using the Vienna
 $ab~initio$ simulation package (VASP), within GGA-PBE ~\cite{PhysRevB.54.11169,prl96Perdew}.
The Kohn-Sham states were expanded in the plane-wave basis set with
 a kinetic energy truncation at 900 eV. In self-consistent calculations, to obtain a converged
  Brillouin zone summation, $9\times9\times9$, $7\times7\times15$, and
  $9\times9\times9~ \boldsymbol{k}$ grids centered at the $\Gamma$ point were 
  applied for $Cmca$-24, $Cmca$-56, and $P4_2/mbc$, respectively. To calculate the
  Fermi surface at a sufficiently dense $\boldsymbol{k}$ 
 grid but at a relatively low computation cost, we employ the Wannier interpolation 
  method as implemented in Ref.~\cite{Mostofi14cpc}. We verified that
  the band structures and density of states obtained with VASP agree
  with those from Abinit.

\section{Results}

\subsection{Crystal structure}
\begin{figure} [t!]
 \centering
\includegraphics[width=0.47\textwidth, clip]{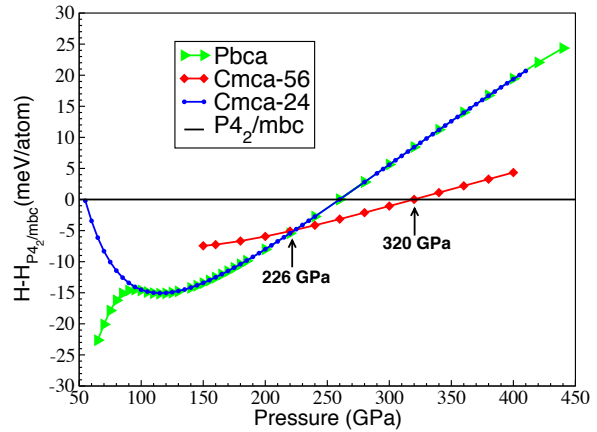}
\caption{ \label{H_HP42} Enthalpies of the most competitive lithium
  structure computed relative to the  {\it P}4$_2$/{\it mbc} structure.}
\end{figure}

The relative enthalpies of the {\it Cmca}-24,  {\it Pbca}, {\it Cmca}-56, and  {\it P}4$_2$/{\it mbc} 
structures in the pressure range from 50 to 500~GPa are shown in Fig.~\ref{H_HP42}. 
These results are consistent with previous
calculations~\cite{PhysRevLett.108.055501,PhysRevLett.102.146401, PhysRevLett.106.015503} and 
experimental data~\cite{Guillaume2011,PhysRevLett.108.055501}.
Space group  {\it Pbca}  is a maximal subgroup of {\it Cmca}. As pressure is
increased, the parameters of the optimized  {\it Pbca}  structure  (space group no. 61; 24-atoms primitive
cell) evolve and by  ~95~GPa it
converges to  {\it Cmca}-24   (space group no. 64; 12-atoms primitive
cell)(see Fig. S15 for x-ray  diffraction comparison of the two phases below and above
95~GPa). The  {\it Cmca}-24 structure (and its equivalent {\it Pbca}) remains
preferred until 226~GPa in agreement with previous calculations~\cite {PhysRevLett.102.146401,PhysRevLett.106.015503}.
Above  226~GPa, {\it Cmca}-56  (28 atom primitive cell) has the lowest enthalpy and remains such
until  about  320~GPa.  At higher pressure, {\it Cmca}-56 becomes
unfavourable with respect to  {\it P}4$_2$/{\it mbc}. While there have
been some disagreements for the exact transition pressures between these
phases~\cite{ PhysRevB.78.014102, PhysRevLett.108.055501,PhysRevLett.102.146401, PhysRevLett.106.015503} they are
 of little consequence for the conclusions of the present
work. The findings reported in what follows for each structure persist
over relatively large pressure ranges. We have therefore selected a
single pressure for each structure, well within their stability
regions, to analyze their electronic properties.

\subsection*{Electronic band structure}
\begin{figure} [t!]
 \centering
\includegraphics[width=0.44\textwidth, clip]{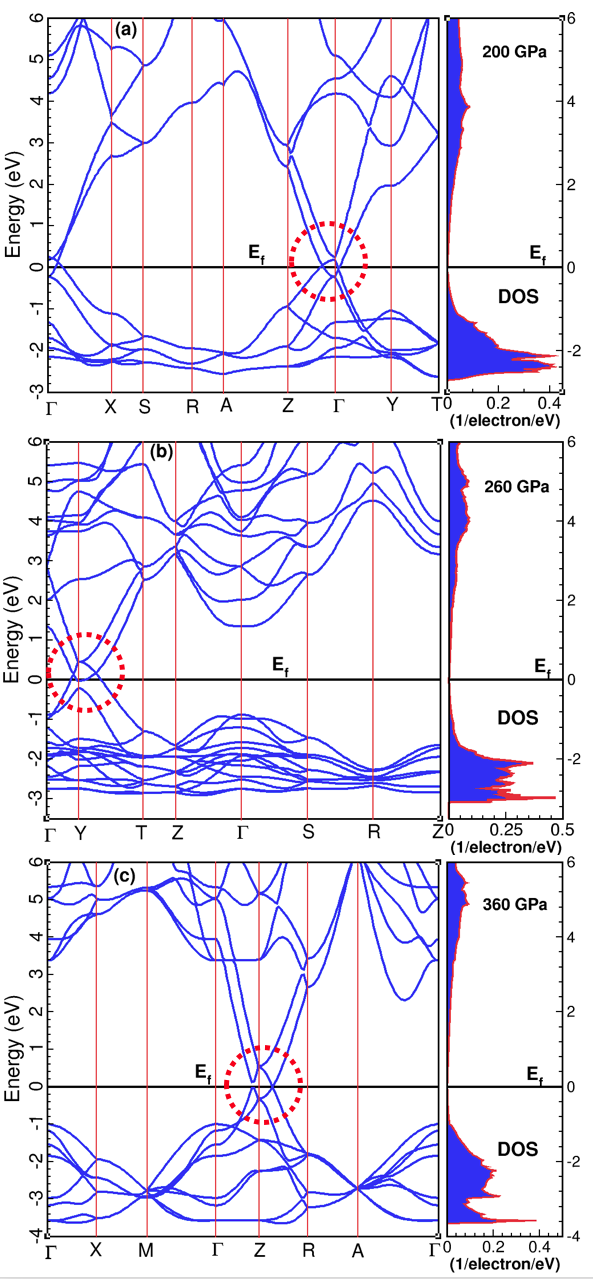}
\caption{ \label{band_dos} Calculated electronic band structures and 
density of states of 
Li (a) {\it Cmca}-24  structure at  200~GPa (b) {\it Cmca}-56  at
260~GPa, and  (c) {\it P}4$_2$/{\it mbc} at  360~GPa.}
\end{figure}

The electronic band structures and densities of states of
the three lithium structures were computed at selected pressures within their
 regions of stability. The results are shown in
Fig.~\ref{band_dos}.
The band structures in all three phases show similar  Dirac-like
characteristics and semimetal behavior. Specifically, the valence and
conduction bands touch each other near the $\Gamma$, $Y$, and $Z$
symmetry points
for {\it Cmca}-24, {\it Cmca}-56, and {\it P}4$_2$/{\it mbc} phases,
respectively. The electronic densities of states of all three structures are
greatly diminished at the Fermi level, however, remain finite. Since
the GGA is known to underestimate the electronic band gap, 
we have carried out hybrid exchange band structures calculations
within the Heyd-Scuseria-Ernzerhof approximation (HSE06) ~\cite{:/content/aip/journal/jcp/124/21/10.1063/1.2204597} as implemented 
 in VASP ~\cite{PhysRevB.47.558,PhysRevB.54.11169}, with otherwise
 exactly the same  simulation parameters as in the GGA-PBE
 calculations. The HSE06 results confirm that the dense lithium phases retain their semimetal behavior [see Fig.~S6(b) for the HSE06
band structure of {\it P}4$_2$/{\it mbc}  at 360~GPa].

\begin{figure} [t!]
 \centering
\includegraphics[width=0.44\textwidth, clip]{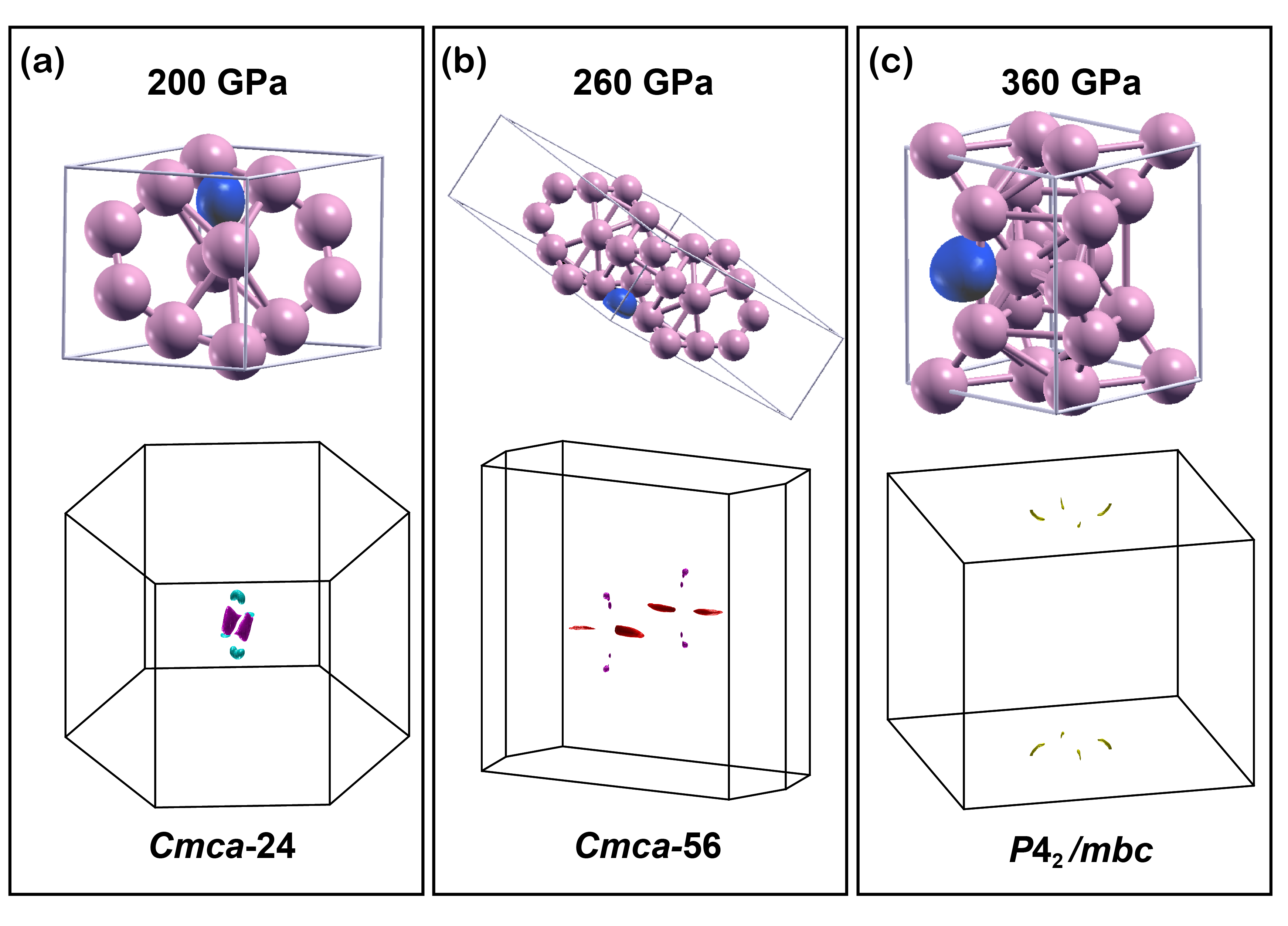}
\caption{ \label{FS_WF} 3D illustrations of the Li crystal structure
  unit cells and Fermi surfaces. The blue balls are 
 isosurface plots of one of the Wannier functions which show electron
 localization in the interstitials. (a) {\it Cmca}-24  phase 
at  200~GPa. (b) {\it Cmca}-56  at  260~GPa, and  (c) {\it P}4$_2$/{\it mbc} at  360~GPa.}
\end{figure}

The semimetal behavior can be further illustrated by
Fermi surface plots, as shown in Fig.~\ref{FS_WF}. Here we see very small Fermi 
surface patches, which indicates that the band crossing points are rather near the Fermi level.

In order to better ascertain the nature of the degeneracies in these high-pressure 
lithium phases, we have constructed Wannier functions for band
 interpolation~\cite{Marzari97prb} for these phases. In Fig.~\ref{FS_WF}, we show
  the Wannier functions of all three phases at selected pressures within their stability 
  ranges; these reveal very peculiar localization of the valence electrons. The Li atoms 
  are depicted as pink spheres. The isosurfaces  of Wannier functions are drawn as blue-colored 
  surfaces. For clarity, only one of the Wannier functions for each structure is drawn
 in Fig.~\ref{FS_WF}. It can be clearly seen that the Wannier wave functions are localized 
 in the interstitials of the rather complex Li networks. The strong interstitial localization is
 again a manifestation the exclusionary effect of ionic cores in compressed
 phases~\cite{Feng08nat}.The localization of electrons can also be seen through electron
 localization functions(see Fig. S10 for details~\cite{SI}). As the electrons are localized in
 the interstitials, these phases can be thought of as elemental electrides,  Li$^+$e$^-$~\cite{Dye03sci}

\subsection*{{Nodal loops}}

From the calculated band structures and density of states shown in
Fig.~\ref{band_dos}, all  three phases are semimetals, with valence
and conduction bands touching at some momenta in the respective Brillouin
zones.
The degeneracies uncovered above may point to topological semimetals, such as Dirac or Weyl semimetals. However, it is noted that the inclusion of spin-orbit coupling in the calculations incurs little changes to the band structure, as expected from the lightness of lithium element. Thus, if these degeneracies assumed Dirac-type linear dispersion, the electronic structures would be correspond to Weyl semimetals. On the other hand, the presence of time-reversal and inversion symmetry in all three phases precludes such possibility, as at least one of the two symmetries must be broken in a Weyl semimetal. These analyses then leave us with the interesting possibility of nodal line semimetals. 

\begin{figure} [tb]
 \centering
\includegraphics[width=0.44\textwidth, clip]{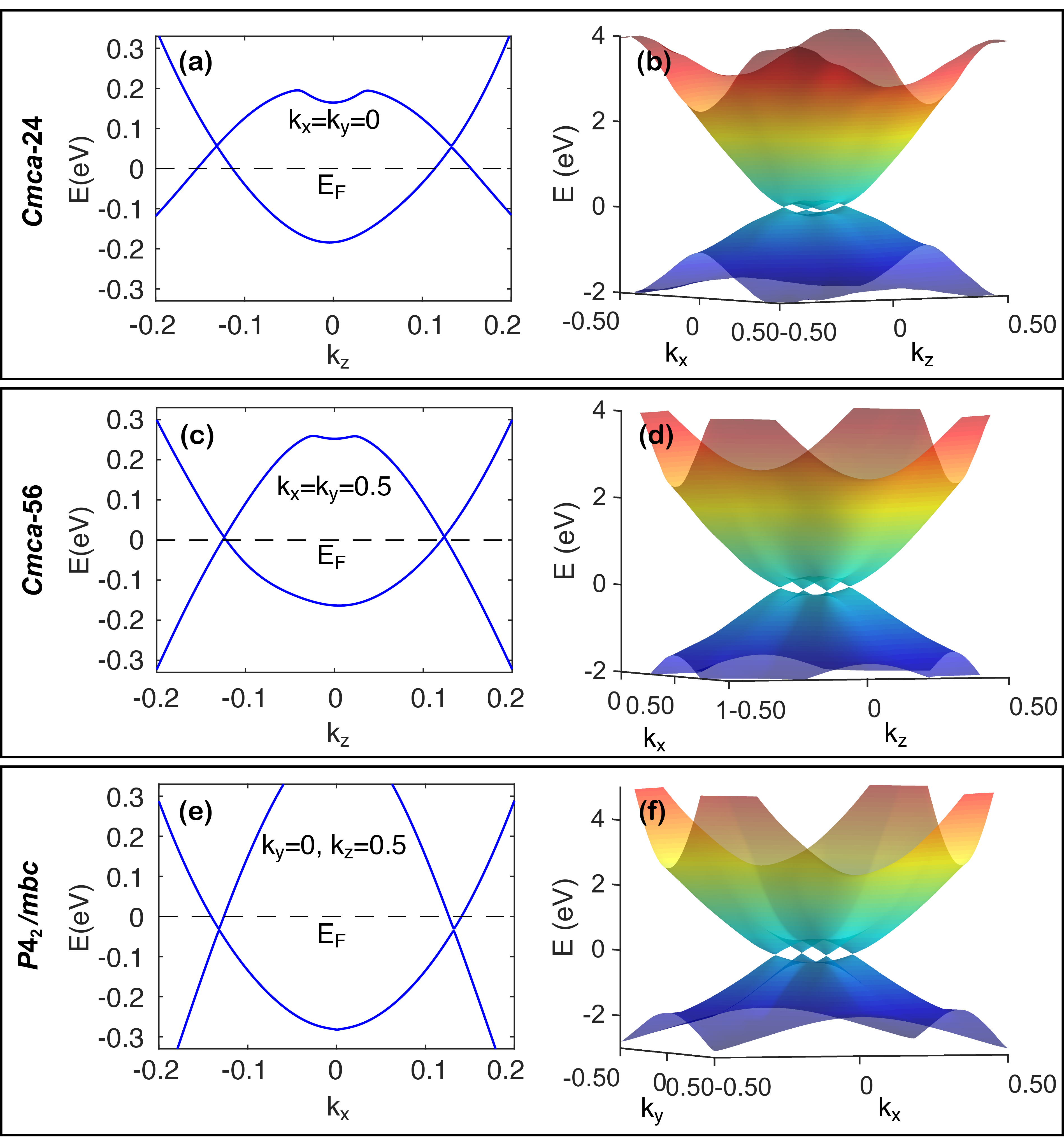}
\caption{ \label{Dirac} Calculated valence and conduction energy bands along
 (a) $k_x=k_y=0$ for the $Cmca$-24 phase at 200 GPa, (c) $k_x=k_y=0.5$ for $Cmca$-56
   at 260~GPa, and (e) $k_y=0, k_z=0.5$ for $P4_2/mbc$  at 360 GPa. (b), (d), and (f) are 
  energy dispersion surfaces in the $k_y=0$ plane for the $Cmca$-24, $k_y=0.5$ for $Cmca$-56, 
  and $k_z=0.5$ for $P4_2/mbc$, respectively.}
\end{figure}

\begin{figure} [tb]
	\centering
	\includegraphics[width=0.44\textwidth, clip]{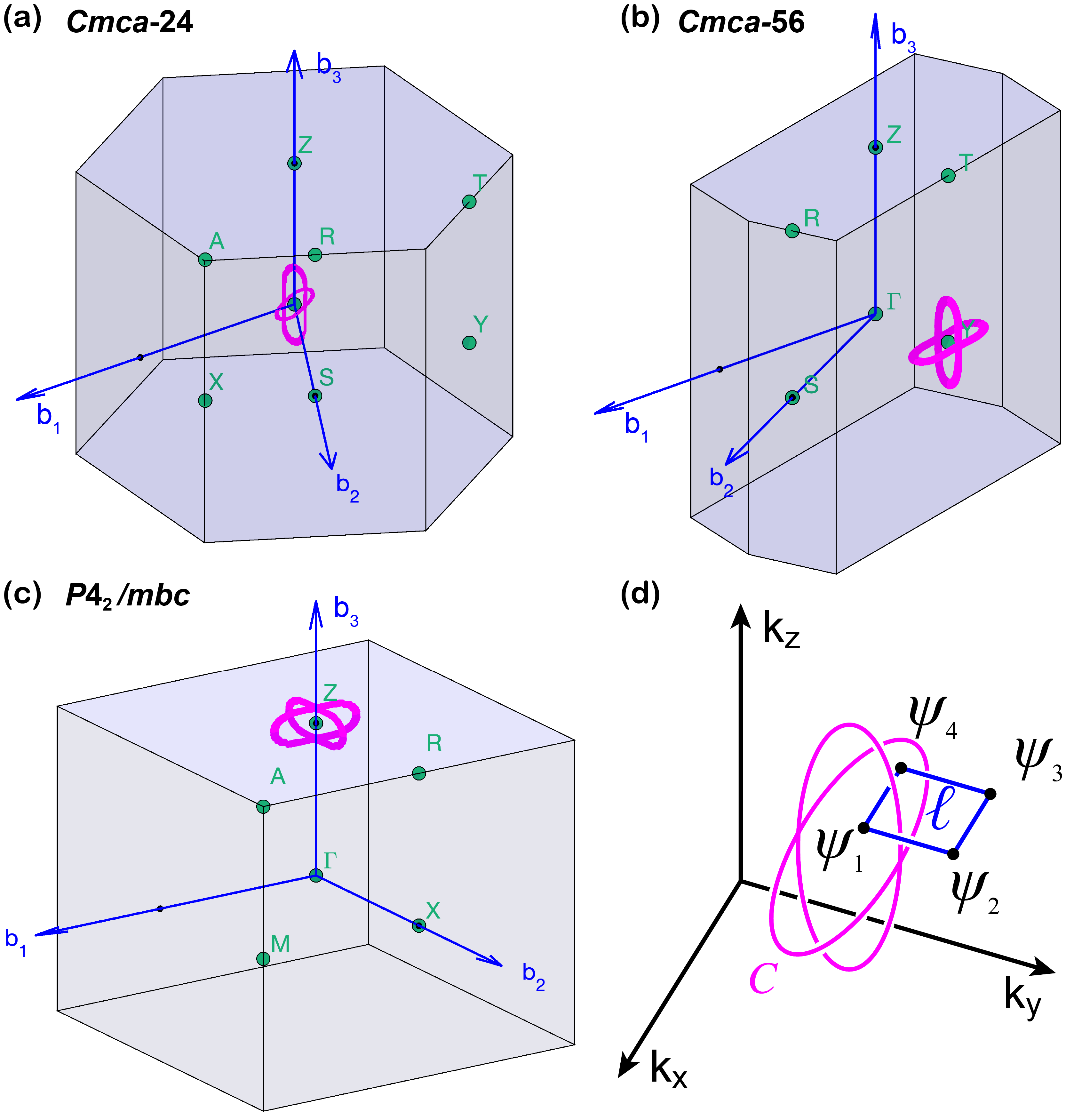}
	\caption{ \label{BZ_DP} Brillouin zones, degenerate nodal loops (magenta curves) for the $Cmca$-24 (a), $Cmca$-56 (b), and $P4_2/mbc$ (c) phases. High-symmetry points are denoted by green points. (d) An example of the closed path $\ell$ (blue rectangle) that encircles the nodal loop $C$ (magenta loop). 
	}
\end{figure}

Therefore, it is essential to scrutinize the band crossings in detail,
which is made possible by expedient band interpolation via Wannier
functions. We plot the dispersions of valence and conduction
bands in two-dimensional $\bm k $-planes and along lines in the
$\boldsymbol{k}$ space. The results are shown in Fig.~\ref{Dirac}. It
can be clearly seen that for the $Cmca$-24 phase there are two Dirac-like
points along the $\boldsymbol{b}_3$ axis, located at $\boldsymbol{k} =
(0,0,\pm0.13)$
 [see Fig.~\ref{BZ_DP}(a)] about 50~meV above the Fermi level. Meanwhile, the energy dispersions in the
$k_y=0$ plane [Fig.~\ref{Dirac}(b)] exhibit four Dirac-like
cones, indicating extra degeneracies. Actually, there are two nodal loops in the $k_z=0$ and $k_x=-k_y$ planes where valence
and conduction bands cross each other [see Figs.~\ref{BZ_DP}(a) and S8~\cite{SI}].
The situation for the $Cmca$-56 phase is similar to that in
$Cmca$-24. There are two Dirac-like points located at the boundary 
of the Brillouin zone with $\boldsymbol{k} =  (0.5,0.5,\pm0.12)$ (Fig.~\ref{BZ_DP}(b)),
almost on the Fermi level, as shown in Fig.~\ref{Dirac}(c). The degenerate nodal loops are
 located in the $k_z=0$ and $k_x=-k_y$ planes [see Figs.~\ref{BZ_DP}(b) and S9~\cite{SI}]. The center of the nodal loops shift to the
  boundary of the Brillouin zone ({\it Y}).
The $P4_2/mbc$ phase,
however, is slightly different. There are two degenerate nodal 
loops in the $k_x=0$ and $k_y=0$ planes [see Figs.~\ref{BZ_DP}(c) and S10 for
details~\cite{SI}]. The center of the nodal loops is located at the boundary of the Brillouin zone ({\it Z}). The 
crossing points between the nodal loops and the $k_z=0.5$ plane are $(\pm0.13,0,0.5)$ and $(0,\pm0.13,0.5)$, as shown in Fig.~\ref{Dirac}(e).
The locations of the degenerate nodal loops within the Brillouin zone of each structure are shown in Fig.~\ref{BZ_DP}.

The nodal loops in all these phases are protected by either the $PT$ symmetry or the glide mirror symmetries. Without spin-orbit coupling, the existence of nodal loop $C$ under $PT$ symmetry requires that the Berry phase for any closed path $\ell$ encircling the loop $C$ is $\pm\pi$~\cite{Zhao16prl}. We pick four $\bm k$-points encircling the candidate nodal lines and compute the Berry phase by summing the Berry connections, as depicted in Fig.~\ref{BZ_DP}(d). The Berry phase is given by 
\begin{align}
\gamma_{\ell}=&\sum_{n\in \mathrm{occ.}} \oint_{\ell}  \langle u_n(\boldsymbol{k})  \vert \mathrm{i} \nabla_{\boldsymbol{k}} u_n(\boldsymbol{k}) \rangle \mathrm{d}\boldsymbol{k}, \nonumber\\
=&-\mathrm{tr}~\mathrm{Im}~  \log \psi_{1}^{\dagger} \psi_{2} \psi_{2}^{\dagger} \psi_{3}\psi_{3}^{\dagger}\psi_{4}\psi_{4}^{\dagger}\psi_{1}
\end{align}
where $\vert u_n(\boldsymbol{k}) \rangle$ is the periodic part of the Bloch eigenstates and the summation is over all occupied states. The second equality is used in numerical evaluation of the spinless Berry phase, in which $\psi_{i}=(\vert u_1(\boldsymbol{k}_i)\rangle, \vert u_2(\boldsymbol{k}_i)\rangle,...,\vert u_M(\boldsymbol{k}_i)\rangle)$, and $M$ is the number of occupied states. A nontrivial Berry phase will protect the nodal loop against weak perturbations from  gap opening. According to our numerical calculations based on tight binding Hamiltonians, all  three of these phases yield nontrivial Berry phases ($\pm\pi$) for closed paths encircling the nodal loops. The presence of glide mirror symmetries will further pin the nodal loops to corresponding planes in the momentum space, as specified in Fig.~\ref{BZ_DP}.

To further examine  and characterize the nature of the degeneracies in these
 high-pressure lithium phases, we construct and analyze an effective $\boldsymbol{k} \cdot \boldsymbol{p}$ Hamiltonian. 
 Here we take the tetragonal phase ($P4_2/mbc$) as an example. We construct a two band model for the valence and
 conduction bands. To capture the degenerate nodal loops
 we expand the Hamiltonian at the $Z = (0,0,1/2)$ point of the Brillouin zone.
Considering the $C_2(z)$ symmetry, we  let $\tilde{k_z}\equiv k_z-0.5$, and the Hamiltonian is written as:
\begin{equation}
\label{Hamiltonian}
	H(\bm k)= \sum_{i=0}^3h_i\sigma^i,
\end{equation}
where $\sigma^i$ are Pauli matrices, and $h_i$ are quadratic functions 
of $\bm k$ determined by the symmetry of $Z$, $i = 0, 1, 2, 3$. Specifically, $h_2 = 0$, $h_i = a_i+b_i
 \tilde{k_z} +c_i k_x^2 +d_i k_y^2 +e_i \tilde{k_z^2}$  for $i=0$ and
 3, and  $h_1(k_x,k_y,\tilde k_z) =f k_x k_y$.
The eigenvalues of Eq.~(\ref{Hamiltonian}) are given by: 
\begin{equation}
\varepsilon_{\pm}(\bm k)=h_0 \pm \sqrt{h_1^2+h_3^2}.
\end{equation}
It can be clearly seen that the valence and conduction bands will
touch each other provided that $h_1=h_3=0$. In the $k_z=0.5$ plane,
this corresponds to four crossing points along the $\boldsymbol{b}_1$ and
$\boldsymbol{b}_2$ directions, i.e., $( q_1,0,0.5)$ and $(0,
q_2,0.5)$, where  $q_1^2 = -a_3/c_3$ and $q_2^2 = -a_3/d_3$. 
In the $k_x=0$ and $k_y=0$ planes, this corresponds to two degenerate
nodal loops. 
The nodal loops are described by elliptic equations. These are $d_3
k_y^2+e_3 \tilde{k_z^2}+b_3 \tilde{k_z}+a_3=0$ in the $k_x=0$ plane and $c_3 k_x^2+e_3 \tilde{k_z^2}+b_3 \tilde{k_z}+a_3=0$ in the $k_y=0$ plane.

The model parameters are fitted from first-principles results and listed in Table ~\ref{fitpara}. 
The locations of crossing points determined by our effective Hamiltonian are $(\pm 0.13,0,0.5)$ and $(0,\pm 0.13,0.5)$, which are consistent with our previous observation.

\begin{figure} [tb]
	\centering
	\includegraphics[width=0.45\textwidth, clip]{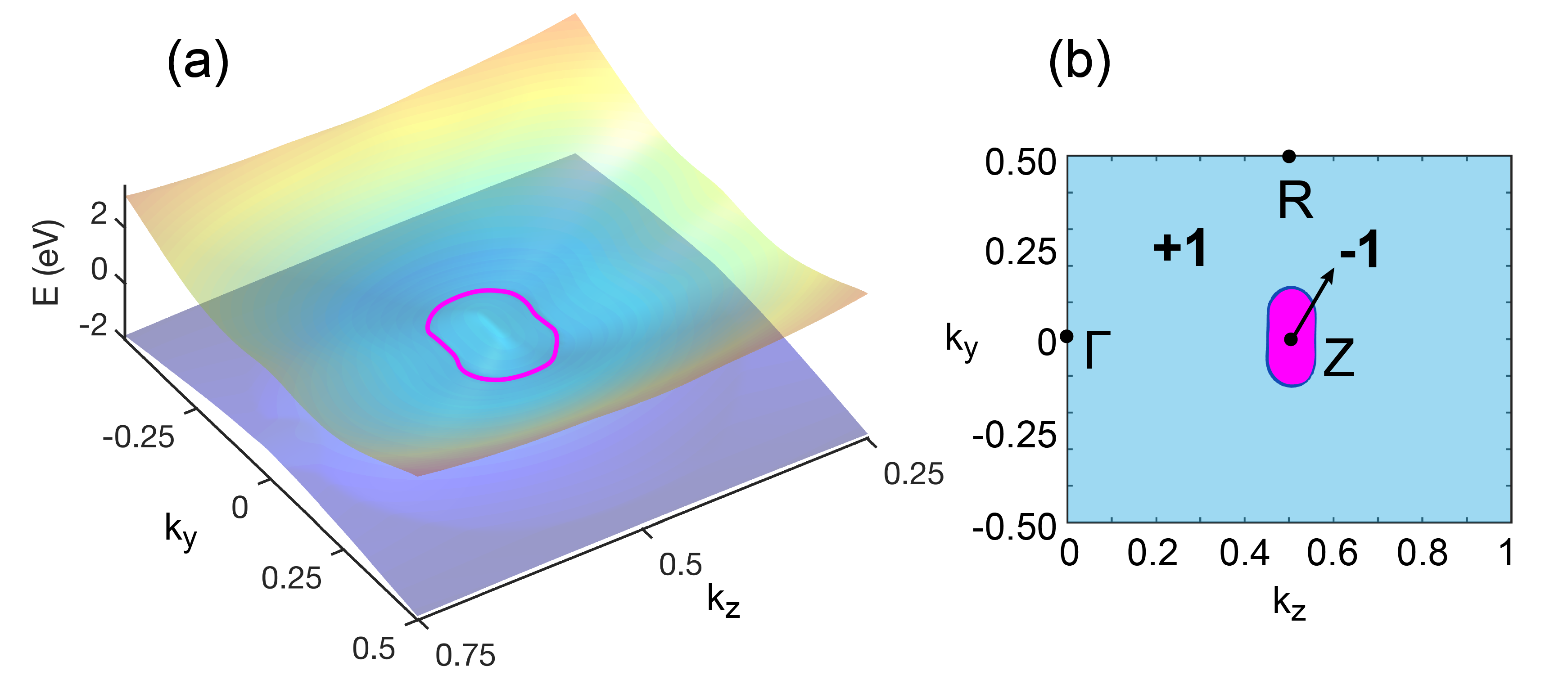}
	\caption{ \label{loop} (a) Energy dispersion of $P4_2/mbc$ phase at the $k_x=0$ plane. The nodal loop is depicted by a magenta line and a dashed line. (b) Eigenvalues of the symmetry operation $c$-glide plane at the $k_x=0$ plane.
	}
\end{figure}

The existence of nodal loops can be further confirmed by symmetry analysis again. Taking the $k_x=0$ plane as an example, the nodal loop can be viewed from both the 3D energy dispersion [Fig.~\ref{loop}(a)] and the two-dimensional Brillouin plane [Fig.~\ref{loop}(b)]. The calculated eigenvalues of the $c$-glide plane are shown in Fig.~\ref{loop}(b). As the areas divided by the nodal loop (magenta) possess opposite eigenvalues with respect to $c$-glide plane with the area outside the nodal loop (blue), there must be band crossings on the boundary. This necessarily leads to degeneracy along nodal lines, which in this case are closed loops.

\begin{table}[htbp]
	\begin{center}
	 \centering
		\caption{\label{fitpara}} Fitted parameters for the
		$\boldsymbol{k} \cdot \boldsymbol{p}$ Hamiltonian for {\it P}4$_2$/{\it mbc} phase in
		Eq.~(\ref{Hamiltonian}) in units of eV.
		\centering
		~\\
		~\\
		\begin{tabular}{p{1.2cm}p{1.2cm}p{1.2cm}p{1.2cm}p{1.2cm}p{1.2cm}}
			\specialrule{0.1em}{1pt}{1pt}
			\hline
			\tc{$a_0 $}   & \tc{$b_0$ }  & \tc{$c_0$}   & \tc{$d_0$}   & \tc{$e_0$}    & \tc{$f$}   \\
			\hline
			\specialrule{0em}{2pt}{2pt}
			\tr{0.053~~~~} & \tr{ 0.002~~~~}  & \tr{-5.106~~~~}    & \tr{-4.921~~~~}   & \tr{-11.210~~~~} &\tr{-89.398~~~}     \\
			\specialrule{0.1em}{1pt}{1pt}
			\tc{$a_3 $}   & \tc{$b_3$ }  & \tc{$c_3$}   & \tc{$d_3$}   & \tc{$e_3$}    &   \\
			\hline
			\specialrule{0em}{2pt}{2pt}
			\tr{-0.317~~~~}   & \tr{-0.210~~~~}    &  \tr{18.082~~~~}  &  \tr{18.531~~~~}  &  \tr{81.830~~~~}  &     \\
			\hline 
			\specialrule{0.1em}{2pt}{2pt}
		\end{tabular}
		
	\end{center}
\end{table}

It should be remarked that recently developed topological quantum chemistry links the crystal symmetries of a given material with its topological properties~\cite{Bradlyn2017}. Our further calculations based on this theory show that the electronic bands of these phases do not satisfy the compatibility relations, thus they are
 indeed to be classified as topological semimetals~\cite{Vergniory2019,Song2019prx}. Interestingly, although the spin-orbit coupling is rather small for light elements such as lithium, the spin-orbit effects will open small gaps (about meV) for the {\it Cmca}-24 and {\it Cmca}-56 phases and turn them into three-dimensional strong topological insulators, while the $P4_2/mbc$ phase remains a topological semimetal.

%

\section*{Conclusions}
In computations on elemental Li at intermediate pressures of 200-360 GPa we find 
an electronic structure characteristic of a semimetal, with a low but non-vanishing 
density of states at the Fermi level. Detailed examination of the three competitive
structural types in this region shows nodal loops in the band 
structure of each, centered at the Brillouin zones or at the boundaries of the Brillouin zones.
The energies of these nodal loops are near the Fermi level. 
All these features can be characterized by
$\boldsymbol{k} \cdot \boldsymbol{p}$ effective Hamiltonians constructed 
by us, based on symmetry analysis of the systems. In this pressure range elemental 
Li is computed to be a topological nodal-loop semimetal.

Nodal-loop semimetals have been predicted to host
  interesting properties, such as anisotropic electron transport and
  density fluctuations~\cite{Mullen15prl,Rhim2016njp}, unusual optical
  response and circular dichroism~\cite{Carbotte2016njp,Liu18prb}, and
  correlation effects and quantum oscillations~\cite{Liu17prb}. Here,
  the nodal loops we predict in lithium under high pressure are rather
  simple and close to the Fermi level. These can serve as potential probes for identifying the predicted topological semimetallic nature of dense lithium  experimentally.

{\it Note added}.Recently, we learned that Mack {\it et al}.~ \cite {2019arXiv190401248M}
  have also found topological features in lithium in the same phases
  and at roughly the same pressure range as in the present work. 

\section*{Acknowledgments}
We thank Max Amsler for discussions and and Prof. Neaton for communicating the work of Mack {\it et al}.~ \cite {2019arXiv190401248M} to us.
This work was supported by the Energy Frontier Research in 
Extreme Environments (EFree) Center, an Energy Frontier Research
Center funded by the U.S. Department of Energy, Office of Science under
Award No. DE-SC0001057, and by LLNL. The work at LLNL was performed
under the auspices of the U.S. DOE under Contract
No. DE-AC52-07NA27344. JF and ZZ are supported by the MOST of the People's Republic of 
China (Grants No. 2018YFA0305601, and No. 2016YFA0301004), NSFC Grant No. 11725415, and by the Strategic
Priority Research Program of Chinese Academy of Sciences, Grant
No. XDB28000000.\\
\textbf{\\ Supplementary Information:\\} 
Supplementary Information accompanies this 
paper at https://journals.aps.org/prmaterials/\\
\textbf{\\Author contributions statement:\\}
S.F.E. and Z.Z. contributed equally to this work.\\
 S.F.E. , S.A.B., and R.H. designed the research. S.F.E.  and Z.Z. performed the DFT
calculations. Z.Z. and J.F. constructed the model Hamiltonian. All authors contributed to the data analysis and manuscript preparation.\\
\textbf{\\Competing financial interests:\\} The authors declare no competing financial interests.\\
\bibliography{ref.bib}

 \end{document}